\documentclass[epj,final]{svjour}
\usepackage{graphicx}
\usepackage{verbatim}
\begin{document}
\title{Tunneling dynamics of correlated bosons in a double
well potential}
\author{Sunayana Dutta$^1$, Apurba Barman$^1$, Anat Siddharth$^1$ \and Ayan Khan\thanks{ayankhan@fen.bilkent.edu.tr}$^2$ \and Saurabh Basu$^1$} %
\authorrunning{S. Dutta et. al.}
\institute{Department of Physics, Indian Institute of Technology
Guwahati, Guwahati, Assam 781039, India \and Department of Physicsical Sciences, IISER-Kolkata, Mohanpur, Nadia, 741252, India} %


\abstract{
The quantum dynamics of a few bosons in a double well potential is
studied using a Bose Hubbard model. 
We consider both signs for the on-site interparticle interaction
and also investigated the situations where they are large and small. Interesting
distinctive features are noted for the tunneling oscillations of these
bosons corresponding to the above scenarios. Further, the sensitivity of the
particle dynamics to the initial conditions has been 
studied. It is found that corresponding to an odd number of
particles, such as three (or five), an initial condition of having
unequal number of particles in the wells has interesting
consequences, which is most discernible when the population
difference between the wells is unity. 
}
\maketitle

\section{Introduction}\label{Sec_1}
Ever since the ultracold bosonic atoms could be experimentally
prepared by confining in a single quantum state, called a
BEC \cite{Anderson,Davis}, the researchers in the field have
intensely tried to broaden their research on cold atoms.
Manipulation of the atomic gas to emulate various quantum many body
phenomena has emerged into an exciting research endeavor of the
atomic physics and the condensed matter physics community which is
aptly complimented by the discovery of optical lattices with
precisely tunable interaction potentials via manipulating the laser
parameters and Feshbach resonance. Such effects are enormously
facilitated by an ultra-clean, phonon-free system, and finally, a
vastly  magnified version of the crystal lattice.

        Tunneling of particles through a classically impenetrable barrier is
a classic problem of quantum physics \cite{marco}. When interaction between these
particles are included, it may help or hinder the tunneling
phenomena. The time resolved tunneling probability may demonstrate
interesting effects of the roles of the interaction parameter(s) and
the initial configuration of the particles. For example, the
phenomena of time evolved pair tunneling of the particles, as
opposed to the individual tunneling across the barrier, can
crucially depend upon the initial state of the system.

        In the regime where the atoms are weakly interacting, tunneling
phenomena of individual particles dominate, as it is the case for
normal Josephson junctions. However as the (repulsive) interaction
grows stronger between the atoms, two of them located at one side of
the barrier cannot tunnel independently and thus a pair tunneling
becomes inevitable. There can also be a `conditional tunneling
regime' where tunneling of a single particle can happen only in the
presence of a second particle that acts as switch \cite{Foling}.

        A simplified version to study correlated particle dynamics in presence
of confining potential is to consider a single particle or a few
particles in a double well potential. Albeit straightforward, it has
the potential to demonstrate a range of fundamental quantum
phenomena with regard to the tunneling dynamics of the particles and
abilities to manipulate them in terms of suppression of the
tunneling probabilities, and thereby trapping them in one of the
wells. Such trapping phenomena are experimentally realized with a
BEC \cite{Levy,Albiez}. The studies involving a few bosons have turned out to be 
more relevant in recent times after experimental successes of the
`Boson Sampling' \cite{crespi,spring,broome} where a small number of bosons 
were used for experimental demonstration of achieving unprecedented control of multiphoton 
interferences in large interferometers.  These techniques offer huge prospects 
of simplifying the quantum computation problem and speeding it up further.

       A two mode approximation, valid when the energy difference between the
two lowest single particle eigenstates is far smaller than all other
energy states, can describe the tunneling between different Bloch
bands in an optical lattice \cite{Wu}. In this work, we shall
investigate a two-site Bose Hubbard model (BHM), which is
the simplest candidate to investigate the dynamics of correlated
bosonic atoms in a double well
potential \cite{Luo,Haroutyunyan,Lignier,Gong,Eckardt} or a bosonic
junction \cite{Gati}. At the outset, it is helpful to mention that
we shall mainly focus on the physics of weak and strong
inter-particle repulsion limits and the sensitivity of initial
conditions on the tunneling dynamics. 

       The dynamical evolution of the Fock space for a simple two-site Bose
Hubbard model (BHM) (without the density exchange term) with two
bosons has been investigated and the tunneling probabilities are
computed as a function of time \cite{Longhi,Smerzi}. However a
detailed analysis of the quantum dynamics in strong and weak
coupling regimes and the sensitivity of the time evolved state to a
variety of initial states were lacking. This is particularly
relevant for engineered waveguide lattices to achieve a certain
preferred final state. Thus it is interesting to consider a few
($N$) bosons ($N>2$, for example, $N=3,4$ etc) as the complexity of
dynamics (compared to $N=2$) is inevitable as the many body effects
will become more conspicuous for a larger assembly of particles.
Furthermore, an inter-site density exchange term in BHM, relevant for
a gas of dipolar bosonic atoms, can also be considered in the
present context. Inclusion of this term in the BHM is known to have
density ordering effects and is responsible for a rich phase
diagram in a lattice\cite{Iskin}.

       Motivated by the above prospects, we have considered a few bosons
in a Bose Hubbard model (BHM) and investigated the quantum dynamics. In particular, we have
investigated the dependencies of the tunneling probabilities in the
strong and weak on-site interaction limits, briefly the effect of the density
ordering term therein and the sensitivity of the dynamics to a
variety of initial conditions. Among other results, the time evolved
dynamics is seen to be crucially dependent on the initial state of
the system in which it is prepared, particularly when the initial
population difference between the two wells is unity for an odd
number of particles. Further, we inlcude a brief discussion on the effect of using an admixture of initial
states on the tunneling oscillations.

       In the following, the presentation of the paper is organized as follows.
The next section deals with studying the quantum dynamics exactly
for a system consisting of a few bosons confined in a
double well potential and described by a BHM on a two site
lattice. Hence, we present our results on the effect of different initial conditions 
on the tunneling dynamics. In particular, we have included a  brief discussion on using
different admixture of states as initial conditions. The implications of our results on the
experimental scenario is presented thereafter.

\section{The Bose-Hubbard Model and the tunneling dynamics for a few bosons}\label{Sec_2}

Even though we are going to restrict ourselves to the
usual (short ranged) Bose Hubbard model \cite{Longhi,Zollner}, we
include an extended density ordering term while deriving the equations of motion (EOM)
with the motivation of investigating its 
competing effects with the on-site term on the
tunneling dynamics. It is relevant to mention that for dipolar
bosons, such extended range interaction potentials are important to
include, as the research of ultracold dipolar gases gained interest
with the experimental realization of Bose condensed $Cr$ atoms which
hosts large long range interactions \cite{Stuhler}. As will be 
immediately clear, the extended term, unlike that for a lattice, only renormalizes the
on-site interaction for a double well.

    For a system of $N$ interacting bosons occupying the weakly coupled 
low lying energy states of a symmetric double well potential, the BHM Hamiltonian is written as,
\begin{eqnarray}
\label{ebhm} \hat{H}_{BHM} & = & -J
(\hat{a}_{1}^{\dagger}\hat{a}_{2}
+\hat{a}_{2}^{\dagger}\hat{a}_{1})+\frac{U}{2}
(\hat{a}_{1}^{\dagger 2}\hat{a}_{1}^{2}+\hat{a}_{2}^{\dagger
2}\hat{a}_{2}^{2})\nonumber\\ +V(\hat{a}_{1}^{\dagger}\hat{a}_{1}
\hat{a}_{2}^{\dagger}\hat{a}_{2}) 
& = & {\hat {H}}_{J} + {\hat {H}}_{U} + {\hat {H}}_{V} ,
\end{eqnarray}

where $\hat{a}_{1}^{\dagger} (\hat{a}_{2})$ are the creation
(annihilation) operators of bosons in the left (right) wells, $J$
being the tunneling parameter $(J> 0)$ between the two modes, $U$ is
the strength of the on-site interaction ($U >0$) and $V$ is the
strength of the extended density interaction (or exchange
interaction) that has, as mentioned earlier, implications in
formation of density order phases and are suitable in the context of dipolar bosons.
It may be noted here that all the energy scales including the time evolution are expressed in units of
tunneling frequency, $J$. It may be noted that the ${\hat {H}_{V}}$ term
includes ${\hat {H}}_{U}$ in the following way \cite{Buonsante},
\begin{equation}
{\hat {H}_{V}} =  V\left [N^{2} - N - \frac{{\hat {H}}_{U}}{U}\right ],
\end{equation}
where $N = n_{1} + n_{2} = a^{\dagger}_{1}a_{1} + a^{\dagger}_{2}a_{2}$. In the
light of this, the density exchange term (${\hat {H}}_{V}$) has no role, apart from
normalizing the interaction strength $U$ to $U'$ ($ = U - V$). This consequently implies a 
noninteracting scenario for $U = V$. The role of the inter-particle interaction 
becomes only relevant for $\ U \neq V$. Further, the negative sign in the expression 
of $U'$ means that it can either be negative (attractive) for $U < V$ \cite{Kolovsky} or positive   
(repulsive) for $U > V$, for which we have considered $V = 2.U$ and $V = 0.5U$ respectively as
the representative values. However due to the symmetric nature of the model, a sign change in
interaction does not contribute in its dynamics. Albeit, in our analysis we have used 
different values corresponding to attractive and repulsive region (as a pathological case) which effectively provides two different interaction strengths manifesting solely the role of interaction magnitude instead of its character. 
Further, we have distinguished the weak and strong coupling
regimes by assuming $U = 0.1$ and $U = 12$ (both in units of the tunneling
frequency, $J$) respectively. The {\it {`strong'}} and {\it {`weak'}}
will carry these values along throughout the manuscript. Other representative
values have been assumed for the computation of tunneling dynamics, however they 
yield no new qualitative inference.

   To obtain the tunneling dynamics, the state vector of the system is expanded in the basis of
Fock states for a constant particle number $N$, as in the following,
\begin{equation}
\label{mdh}
|\Psi(t)\rangle=\sum_{l=0}^{N}\frac{c_{l}(t)}{\sqrt{l!(N-l)!}}
\widehat{a}_{1}^{\dag l}\widehat{a}_{2}^{\dag N-l}| 0 \rangle,
\end{equation}

where out of $N$ particles, $l$ are in the left well and $N-l$ are
in the right well and $c_{l}(t)$s are the complex coefficients.

    The EOM can be written as,
\begin{equation}
\label{ndh}
i\hbar \frac{d|\Psi(t)\rangle}{dt} =\hat{H}_{BHM}
|\Psi(t)\rangle.
\end{equation}

    In terms of the coefficients $c_{l}(t)$, EOM is expressed as,

\begin{equation}
\label{dis}
i\frac{dc_{l}(t)}{dt}= - k_{l}c_{l+1}- k_{l-1}c_{l-1}+
a_{l}c_{l} + b_{l}c_{l},
\end{equation}
where $k_{l} = J \sqrt{(l+1)(N-l)}$,
$a_{l}=\frac{U}{2}[l^{2}+(N-l)^{2}-N]$, $b_{l}= V[l(N-l)]$

    For the case two bosons ($N=2$), Eq.(\ref{dis}) reduces to
 three coupled equations as in the following,
\begin{eqnarray}\label{dfl}
i\frac{d}{dt}\Biggl(\begin{array}{c}
c_{0}\\
c_{1}\\
c_{2}\end{array}\Biggr)&=&\Biggl(\begin{array}{c c c}
U & -\sqrt{2}J & 0\\
-\sqrt{2}J & V & -\sqrt{2}J\\
0 & -\sqrt{2}J & U \end{array}\Biggr)\Biggl(\begin{array}{c}
c_{0}\\
c_{1}\\
c_{2}\end{array}\Biggr).
\end{eqnarray}
Hence the particle occupation probabilities can be obtained by
solving these coupled equations. The various initial conditions that
can be thought of are,
\begin{eqnarray}
\label{mmm}
c_{0}(0)&=&1, c_{1}(0) = c_{2}(0) = 0;\quad 
c_{1}(0) = 1, c_{0}(0) = c_{2}(0) = 0;\nonumber\\
c_{2}(0)&=&1, c_{0}(0) = c_{1}(0) = 0. 
\end{eqnarray}
In short we shall denote them as $(100), (010)$ and $(001)$
respectively, where $(100)$ means that, initially all the bosons are
in the right well with the left one being empty. Similarly, $(010)$
denotes one in each well, while $(001)$ implies both in the left
well with the right well being empty.

    Similarly, for the case of three bosons, one gets four coupled
equations, which are,
\begin{eqnarray}\label{dfl1}
i\frac{d}{dt}\left(\begin{array}{c}
c_{0}\\
c_{1}\\
c_{2}\\
c_{3}\end{array}\right)&=&\left(\begin{array}{c c c c}
3U & -\sqrt{3}J & 0 & 0\\
-\sqrt{3}J & U+2V & -2J & 0\\
0 & -2J & U+2V & -\sqrt{3}J\\
0 & 0 & -\sqrt{3}J & 3U \end{array}\right)\left(\begin{array}{c}
c_{0}\\
c_{1}\\
c_{2}\\
c_{3}\end{array}\right).\nonumber\\
\end{eqnarray}
As earlier, these equations can be solved for four different initial
conditions, namely, $(1000), (0100), (0010)$ and
$(0001)$, with implications as before. For example, $(0100)$ denotes
a situation where two particles are in the right well, with the
other in the left well and so on.

      A straightforward extension yields similar set of equations
(now a set of five) and the corresponding initial
conditions for $N= 4$. We have repeated the procedure till $N = 16$.
For brevity, we skip them here.

An extension of our results to the case of $N$ bosons is possible
via the method of induction. A straightforward application of this
method on Eqs.(\ref{dfl}) and (\ref{dfl1}) yileds,
\begin{eqnarray}\label{eq:1}
i\frac{dc_{0}}{dt}&=&-\sqrt{N}Jc_{1}+\frac{N(N-1)}{2}Uc_{0},
\end{eqnarray}
where all the bosons are in the right well. Similarly the case
corresponding to $(N-1)$ bosons in the right well and one in the
left, can be denoted by,
\begin{eqnarray*}\label{eq:2}
i\frac{dc_{1}}{dt}&=&-\sqrt{N}Jc_{0}-\sqrt{2(N-1)}Jc_{2}\nonumber\\
&+&\frac{N^2-3N+2}{2}Uc_{1}+(N-1)Vc_{1}.
\end{eqnarray*}
For an equal distribution of the bosons with $N/2$ in each well ($N$: even) is
\begin{eqnarray*}\label{eq:3}
i\frac{dc_{N/2}}{dt}&=&-\frac{J}{2}\sqrt{N(N+2)}\left(c_{N/2+1}+c_{N/2-1}\right)\nonumber\\
&+&\frac{U}{4}N(N-2)c_{N/2}+\frac{V}{4}N^2c_{N/2}.
\end{eqnarray*}
However for discussing our results in the following section, we restrict
ourselves to the case of a few bosons.

\section{Physical Observables and Results} \label{Sec_3}

As emphasized earlier, we are interested in studying the quantum
tunneling dynamics of a few bosons in a double well potential. We
are mainly interested in the effect of the inter-particle repulsion
and density exchange for two, three, four and five bosons and the
variation in the tunneling dynamics associated with different
initial conditions. In the following, we describe the cases
corresponding to two, three and four bosons separately. In this
regard, an useful (and experimentally measurable) quantity to study
the tunneling dynamics of bosons can be the population in one of the
wells (say the right well), $P_{R}(t)$ as discussed in the following discussion \cite{Zollner}.

   For a system of two bosons, the right well population, $P_{R}(t)$ can be
expressed as \cite{Longhi},
\begin{equation}
\label{ooo}
 P_{R}(t) = |c_{0}(t)|^{2} + \frac{1}{2}|c_{1}(t)|^{2},
\end{equation}
which is a superposition of the probabilities of both the bosons in the right well
and half of that corresponding to one in each well.

  It may be noted that for $U' = 0$ ({\it {i.e.}} $U = V$), Rabi oscillation of the particles
between the wells is observed (Fig.1(a)). At small values of $U'$, the atoms can
still tunnel independently, similar to that of the normal Josephson
junction, however the time period for oscillation becomes enormously
large which signals the onset of a trapping scenario. A similar
scenario has been reported by Z\"{o}llner $et$ $al.$ \cite{Zollner},
where they have found that for $g = 1.3$ ($g$ being the strength of
the pairwise potential), the time period is as large as $2 \times
10^{3} ~ s$. In this situation, two or more atoms residing in one of
the wells, form a `repulsively bound pair' \cite{Winkler} and hence
tunnel together. Such phenomena are difficult to contemplate in
crystal lattices owing to relatively much shorter life times
associated with the decay processes.

      For the weakly interacting case $U = 0.1$, it is important to note 
that $P_{R}(t)$ collapses, as evident from Fig.1(b). However there is again a `revival' as 
time progresses and this phenomenon is repeated with increasing time. At time, $t=0$, the
system is prepared in a definite state (as described by the initial
conditions in the preceding section) and the two terms in
Eq.(\ref{ooo}) are correlated. However as time increases, the
oscillations corresponding to different initial excitations pick up
different frequencies and hence become uncorrelated, thereby leading
to a collapse. With further increase in time, the correlation is
partly (depending on the value of $U$) restored and revival occurs.
This behavior repeats itself and thus an infinite sequence of
collapse and revivals are obtained \cite{Scully}.
\begin{figure*}[!t]

\includegraphics[width=5.5cm]{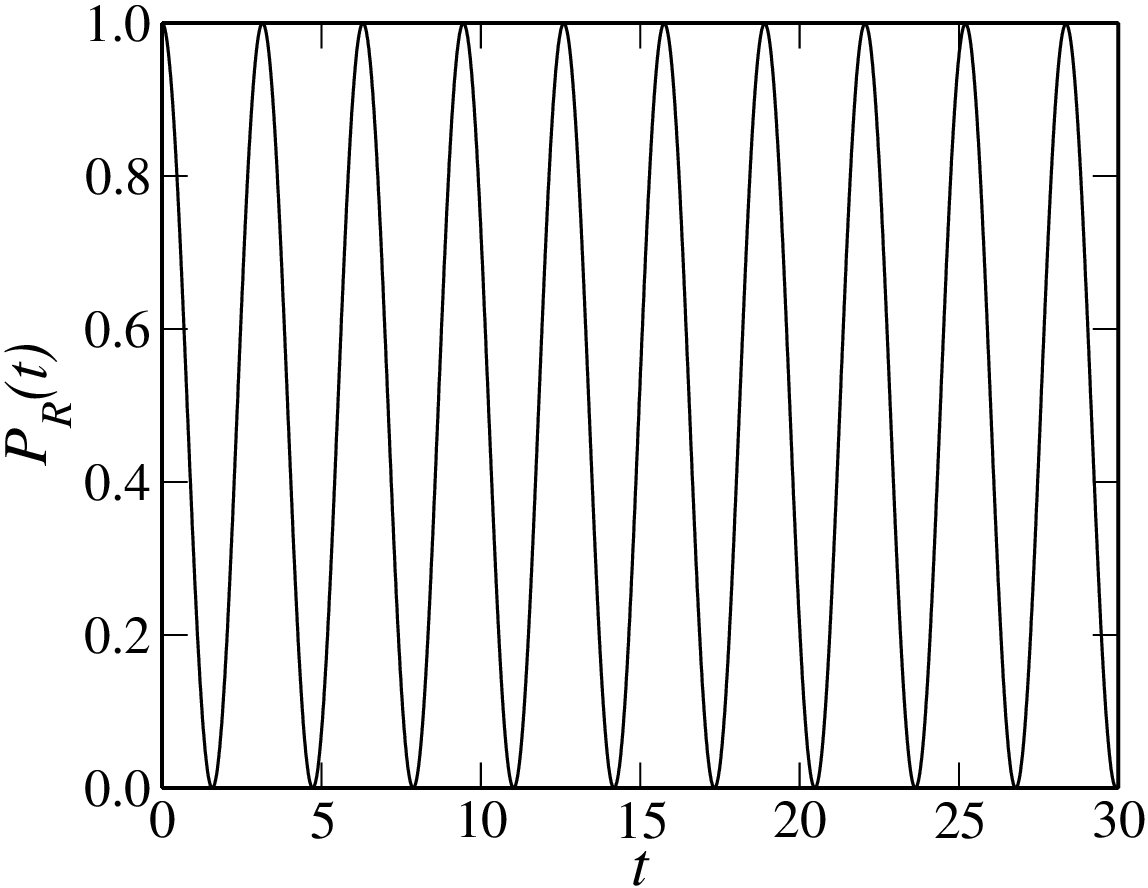}
\includegraphics[width=5.5cm]{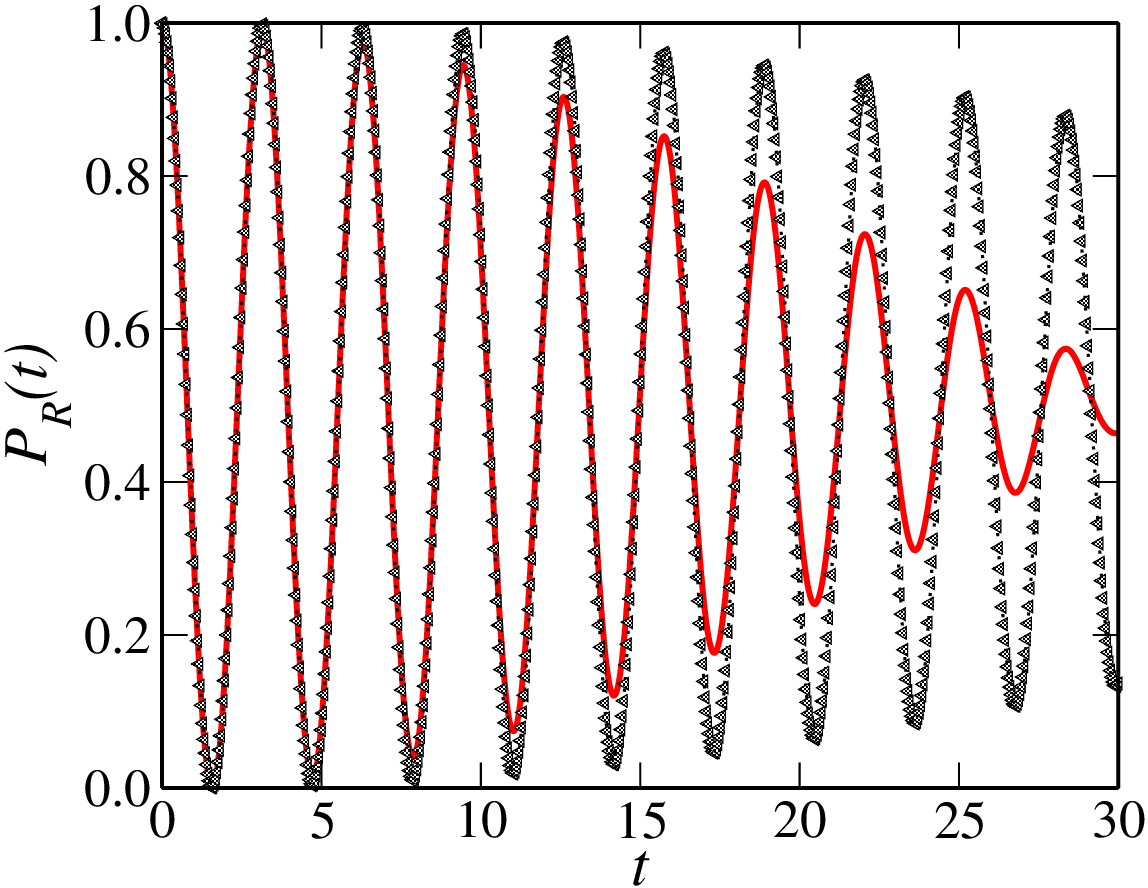}
\includegraphics[width=5.5cm]{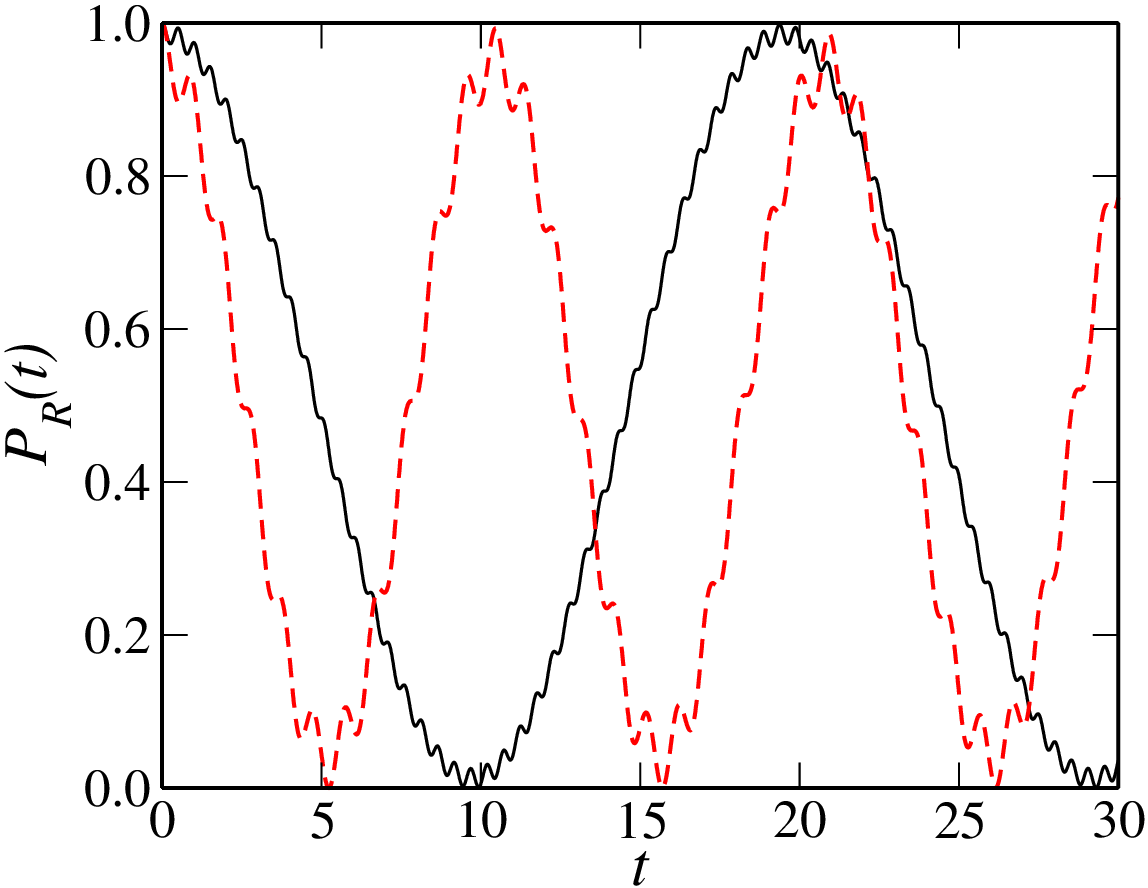}
\centerline{\qquad$(a)$ \hfill\qquad  $(b)$ \hfill\qquad $(c)$ \hfill}
\caption{\label {Interaction}(Color online)  $P_{R}(t)$, defined in text,
as a function of time for two bosons is
shown for non interacting case ($V = U$) in $(a)$. In $(b)$ 
 left triangles with dots (black) denote $P_{R}(t)$ corresponding to $U'=0.05$ and 
solid line (red) denotes $P_{R}(t)$ corresponding to $U'=-0.1$. Similarly in $(c)$ 
dashed lines (red) denote $P_{R}(t)$ corresponding to $U'=6$ and solid
line (black) denotes $P_{R}(t)$ corresponding to $U'=-12$ The time $t$ in the $x$-axis is
measured in units of the tunneling frequency, $J$ and is true for all plots.}
\end{figure*}
   At large values of $U$, namely, $U=12$, termed as the {\it
{fermionization limit}} (where bosons avoid each other and thus obey
an \textit{`exclusion principle'}), there are faster oscillations
with smaller amplitudes, however $P_{R}(t)$ becomes zero eventually (see Fig.1(c)),
signaling a tunneling of the atoms at large time scales. The time
period of such `eventual tunneling' phenomena, $T_{R}$ (say)
increases with the increase in the inter-particle repulsion, $U$,
thereby signaling intense trapping effects. The analytic expressions for $T_{R}$
corresponding to the noninteracting ($U = V$), and interacting
(considering two pathological cases $V = 2.U$ and $V = 0.5U$) cases are obtained as,
\begin{eqnarray}
T_{R} & = &  \pi/J~~~~{\rm {for}}\,\,V = U \\ \nonumber
              & = & \left|\frac{8\pi}{U - \sqrt{64J^{2} + U^{2}}}\right| ~~~~{\rm {for}}\,\,V = 0.5U \\ \nonumber
              & = & \left|\frac{4\pi}{U - \sqrt{16J^{2} + U^{2}}}\right| ~~~~{\rm {for}}\,\,V = 2U
\end{eqnarray}
Thus, as a function of $U$, $T_{R}$ scales linearly and the agreement between the analytic expressions and
the corresponding numeric estimates are shown in Fig. (2). So $T_{R}$ is insensitive to $U$ values
for the non-interacting case, while it sharply increases as $U$ is increased 
for the interacting cases.
\begin{figure*}[!t]
\centering
\includegraphics[width=7.5cm]{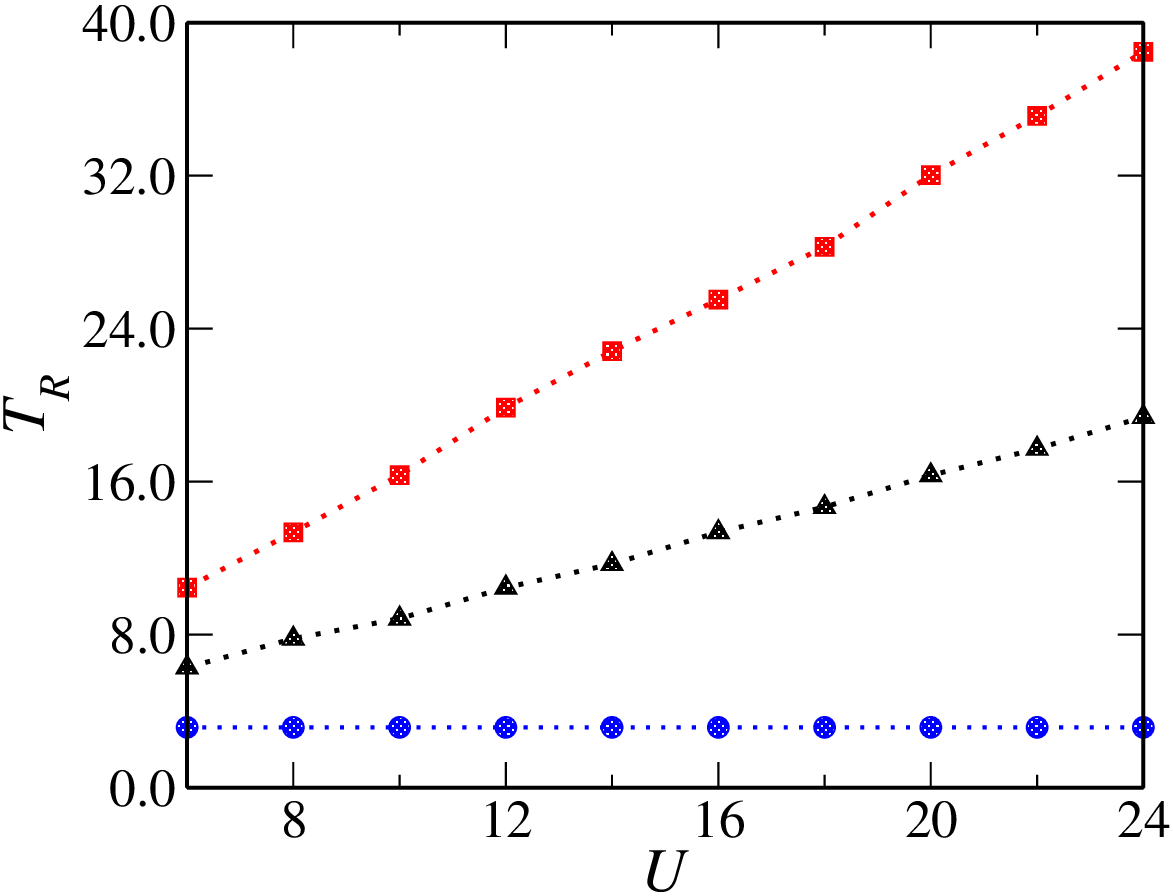}
\caption{\label {Time period}(Color online) Time period $T_{R}$ of two bosons as a
function of $U$ is shown. In the figure, the circles with dotted line (blue), triangles with dotted line (black)
and squares with dotted line (red) denote the analytical results for the non-interacting case  ($V = U$),
and interacting cases with $V=0.5U$ and $V=2U$ respectively. The associated open squares are the 
numerically obtained results.}
\end{figure*}

    Let us now analyze the impact of different initial conditions on the
tunneling dynamics of two particles in a double well potential. To
prepare an initial state with a population imbalance, a tilt in the
form of a linear potential $- \eta x$ ($\eta > 0$) can be
superimposed \cite{Zollner,Sachdev}. For a reasonably large $\eta$
(magnitude of the tilt), all the particles can be made to reside in
one well. The subsequent dynamics can be studied by allowing $\eta
\rightarrow 0$ within some characteristic time scale. Motivated by
such prospects of experimentally creating different initial
states \cite{Frazer}, we study the tunneling dynamics subject to
different initial conditions. 

In the weak coupling regime ($U = 0.1$) and the so called attractive limit
($U' < 0$ or $V = 2.U$), $P_{R}(t)$ oscillates and slowly dampens for
the initial conditions given by (100) and (001) (see Fig.3(a)), while
the damping is faster with further weakening of the interaction field ($U' > 0$ or $V = 0.5U$) (Fig.3(b)). The
situation in the strong coupling limit ($U = 12$) show Rabi oscillations with different
frequencies corresponding to the $V=2U$ (Fig.3(c)) and $V=0.5U$ (Fig.3(d)) situations.
It may also be noted that if we start with an initial condition $(010)$ where one boson resides in each well, Eq.$(3)$ becomes,

\begin{equation}
|\Psi(t)\rangle={c_{1}(t)}
\widehat{a}_{1}^{\dag }\widehat{a}_{2}^{\dag }| 0 \rangle  \nonumber
\end{equation}
It can be shown that the above state is an eigenstate of the Hamiltonian (Eq.$(1)$) and hence the dynamics is frozen which is seen from Fig.$(3)$, where $P_R(t)$ 
stays at 0.5 irrespective of the values of parameters used.
The frequency and time period of these oscillations depend upon the interaction parameters used in this
work.
\begin{figure*}[!t]

\includegraphics[width=4.25cm]{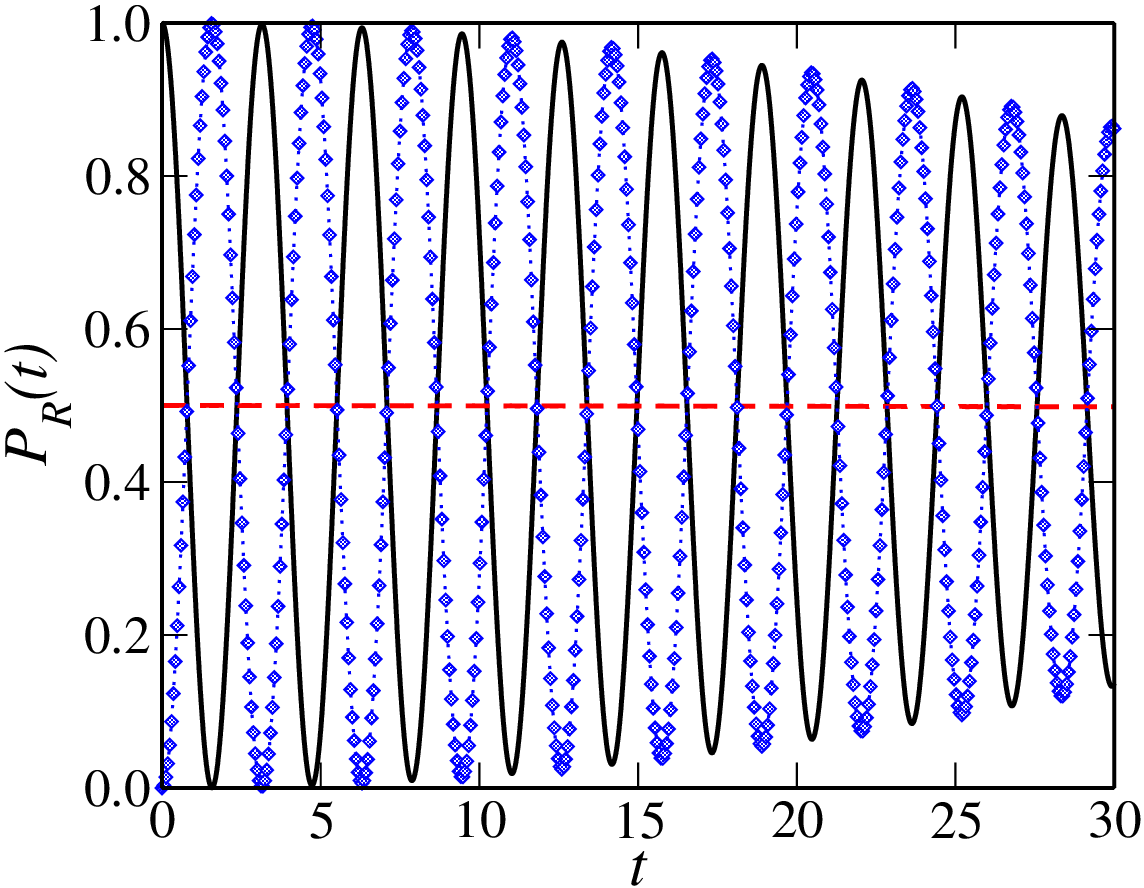}
\includegraphics[width=4.25cm]{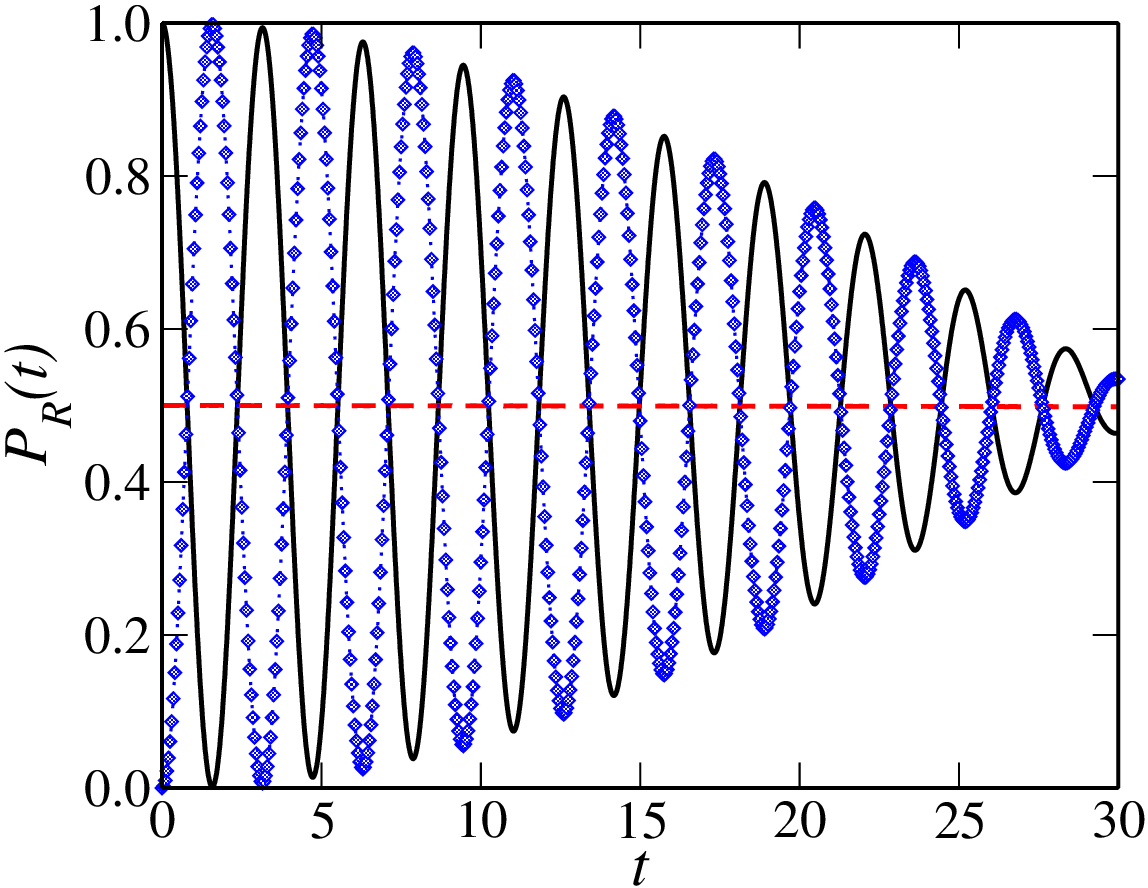}
\includegraphics[width=4.25cm]{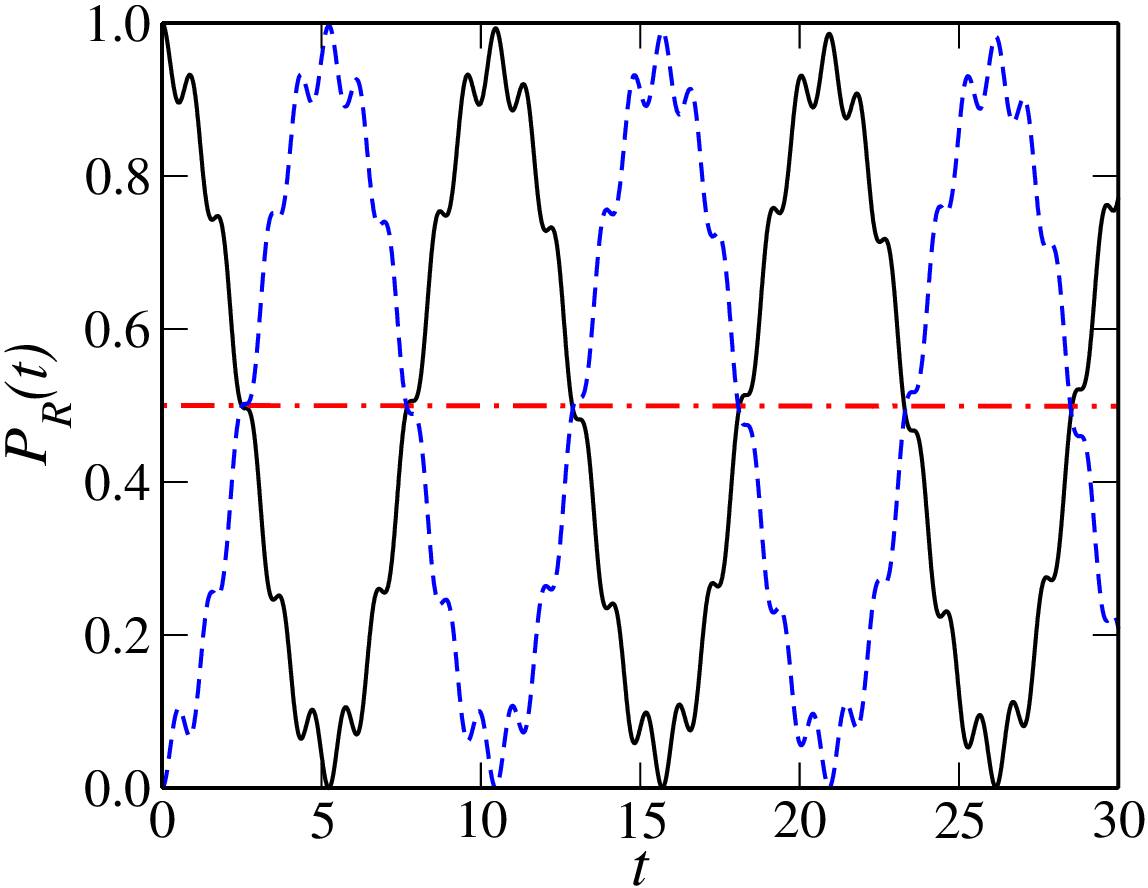}
\includegraphics[width=4.25cm]{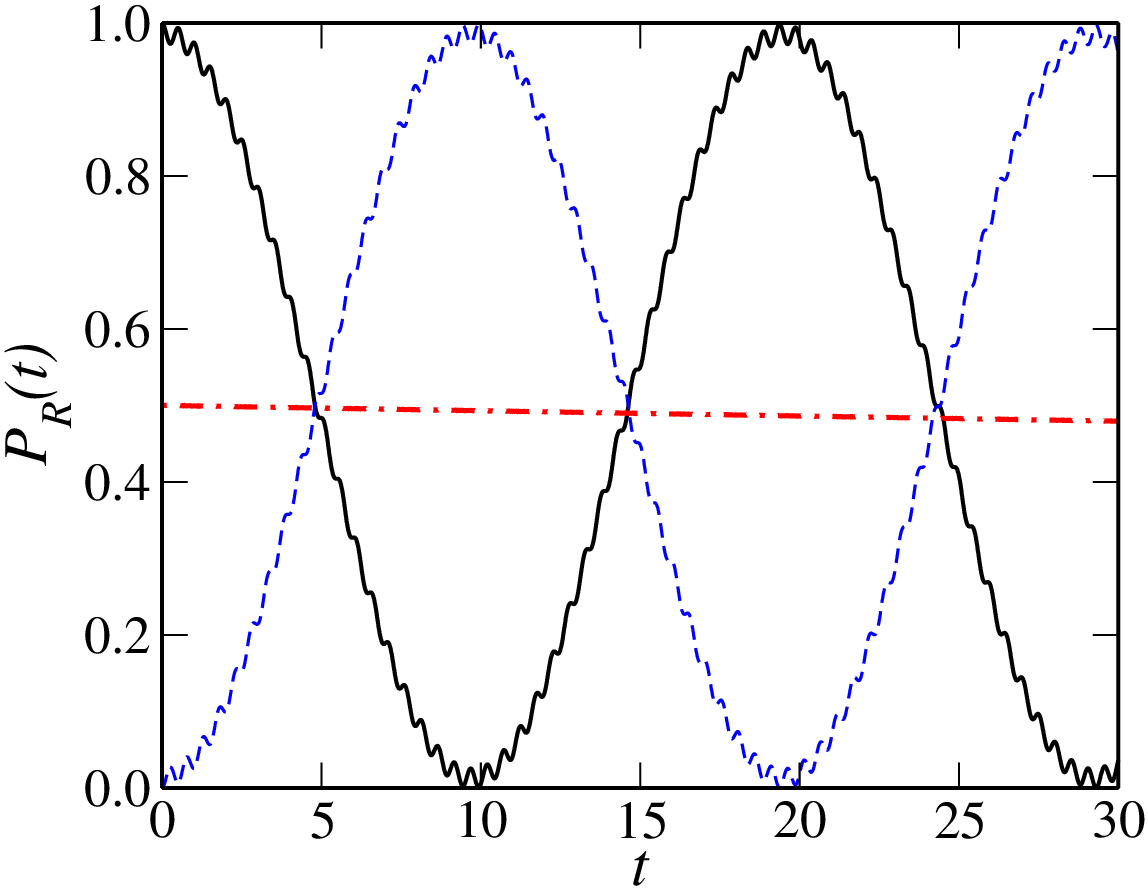}
\centerline{$(a)$ \hfill\hfill $(b)$ \hfill\hfill $(c)$ \hfill\hfill  $(d)$ \hfill\hfill}
\caption{\label {Initial Condition}(Color online)  $P_{R}(t)$ as a
function of time corresponding to three different initial
conditions is shown. In $(a)$ $U'=0.05$, and $(b)$ $U'=-0.1$ solid line (black) denotes 
$(100)$, dashed line (red) denotes $(010)$  and diamonds with dots (blue) denote to$(001)$.
 In $(c)$ $U'=6$ and $(d)$ $U'=-12$  solid line (black) denotes $(100)$, dashed-dotted line
(red) denotes $(010)$, and dashed lines (blue) denote $(001)$. See text for details on notations.}
\end{figure*}

       Let us now concentrate on the case of three bosons ($N=3$).
The $P_{R}(t)$ in this case is defined as,
\begin{equation}\label{hhh}
P_{R}(t) = |c_{0}(t)|^{2} + \frac{2}{3}|c_{1}(t)|^{2} +
\frac{1}{3}|c_{2}(t)|^{2}.
\end{equation}
Similar to the case of two bosons, here $P_{R}(t)$ is a combination of
probabilities of all of them in the right well ($|c_{0}(t)|^{2}$),
two in the right and one in the left with an amplitude $\frac{2}{3}$
and one in the right and two in the left with an amplitude
$\frac{1}{3}$ respectively. While qualitatively the tunneling behavior remains unaltered as compared to two bosons, with regard to Rabi oscillations at $U'=0$ (not shown here) and there is a temporary decay of the amplitude of oscillations (Fig.$4(a)$) due to the beating phenomena for $U'\ne0$. It can also be seen that $V=0.5U$ registers more significant decay of the amplitude owing to trapping effects. At large $U$, (Fig.$4(b)$), the time period of {\it {`eventual'}}
oscillations becomes very large. The time period is
about an order of magnitude larger compared to two bosons. In Fig.4(b), for $U'=6$, one can observe
tunneling phenomena at larger time scales, while at $U'=-12$ (large $V$), the tunneling of atoms
take a very long time and we do not observe any tunneling until $t=60$ and even to much large values of time (not shown here).
This indicates emergence of trapping phenomena for large values of the
exchange intercation $V$ in the large $U$ regime.

\begin{figure*}

\includegraphics[width=4.25cm]{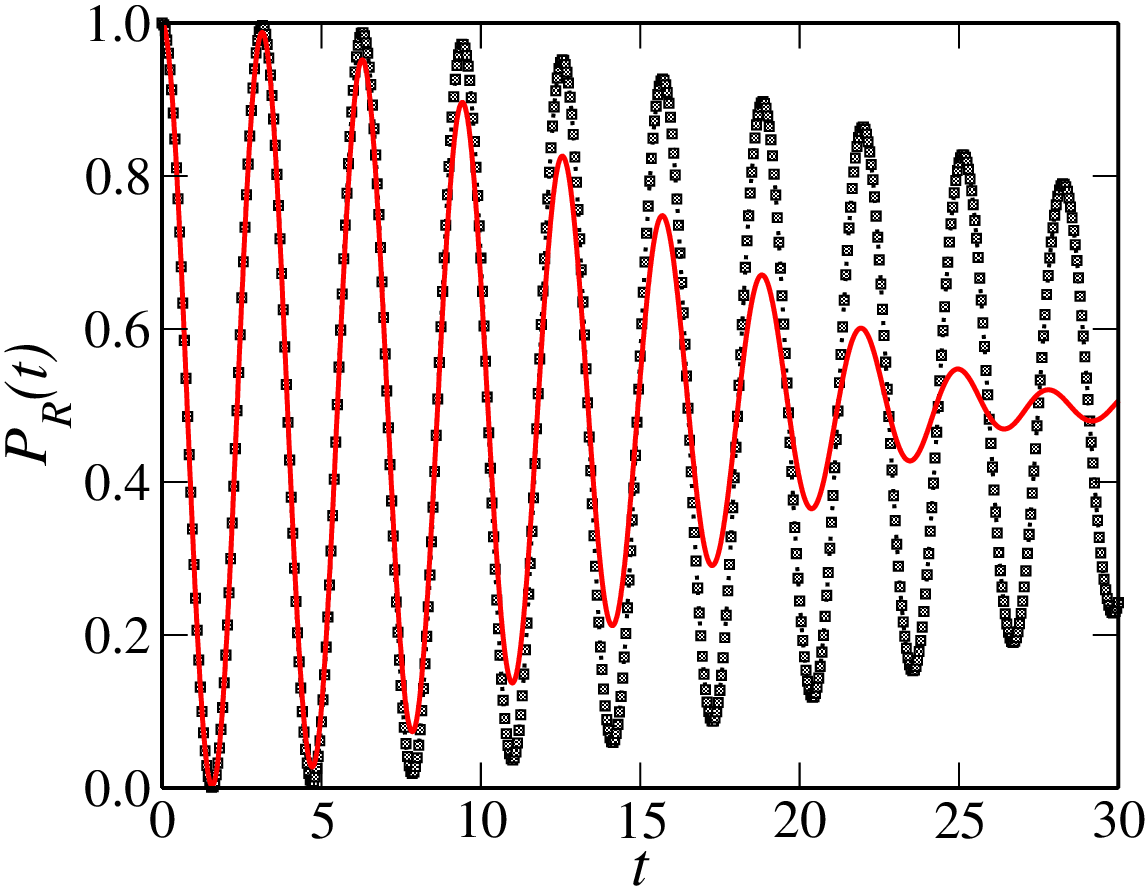}
\includegraphics[width=4.25cm]{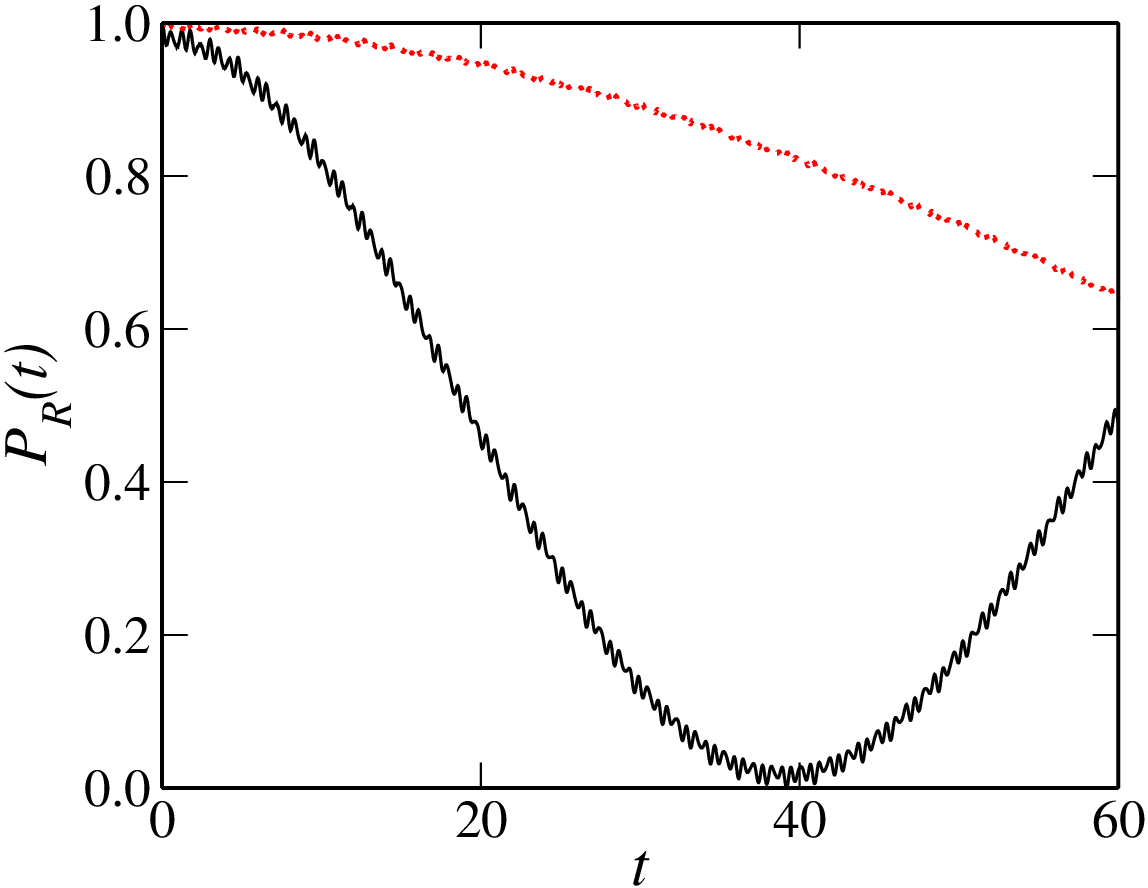}
\includegraphics[width=4.25cm]{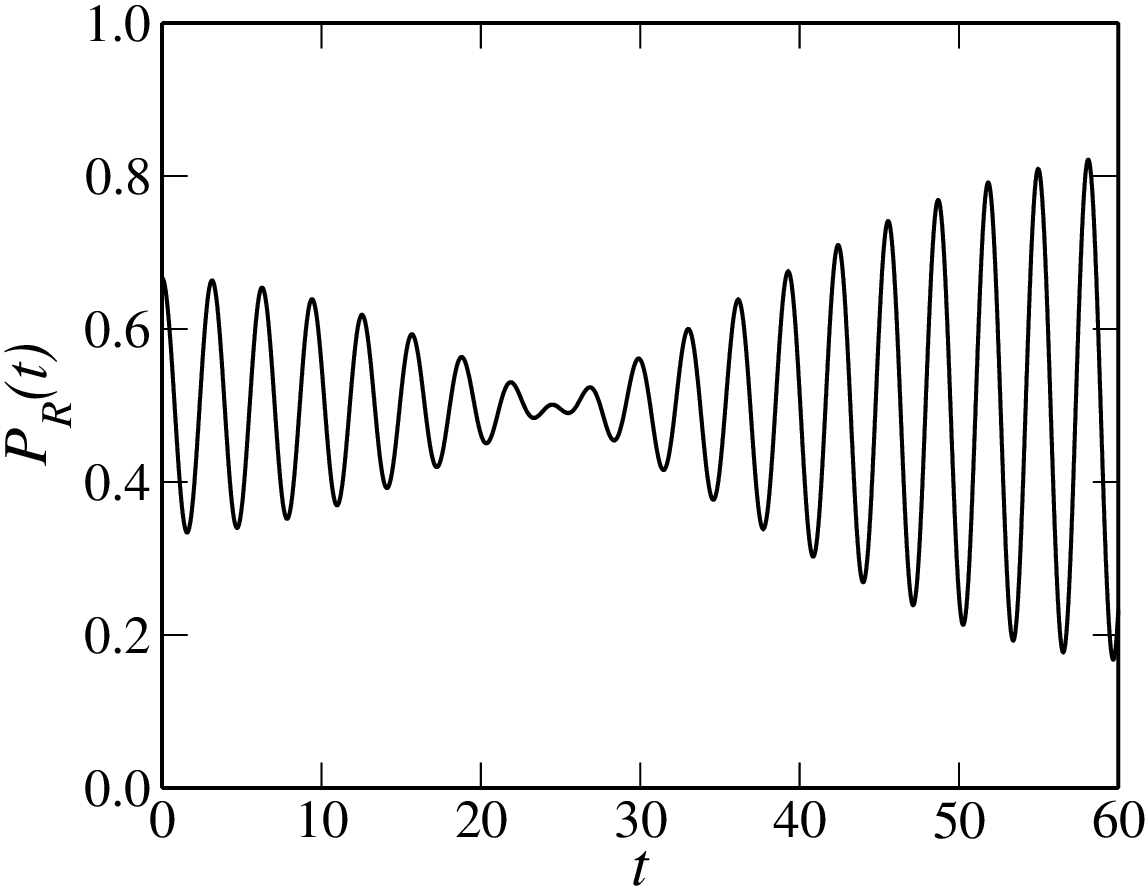}
\includegraphics[width=4.25cm]{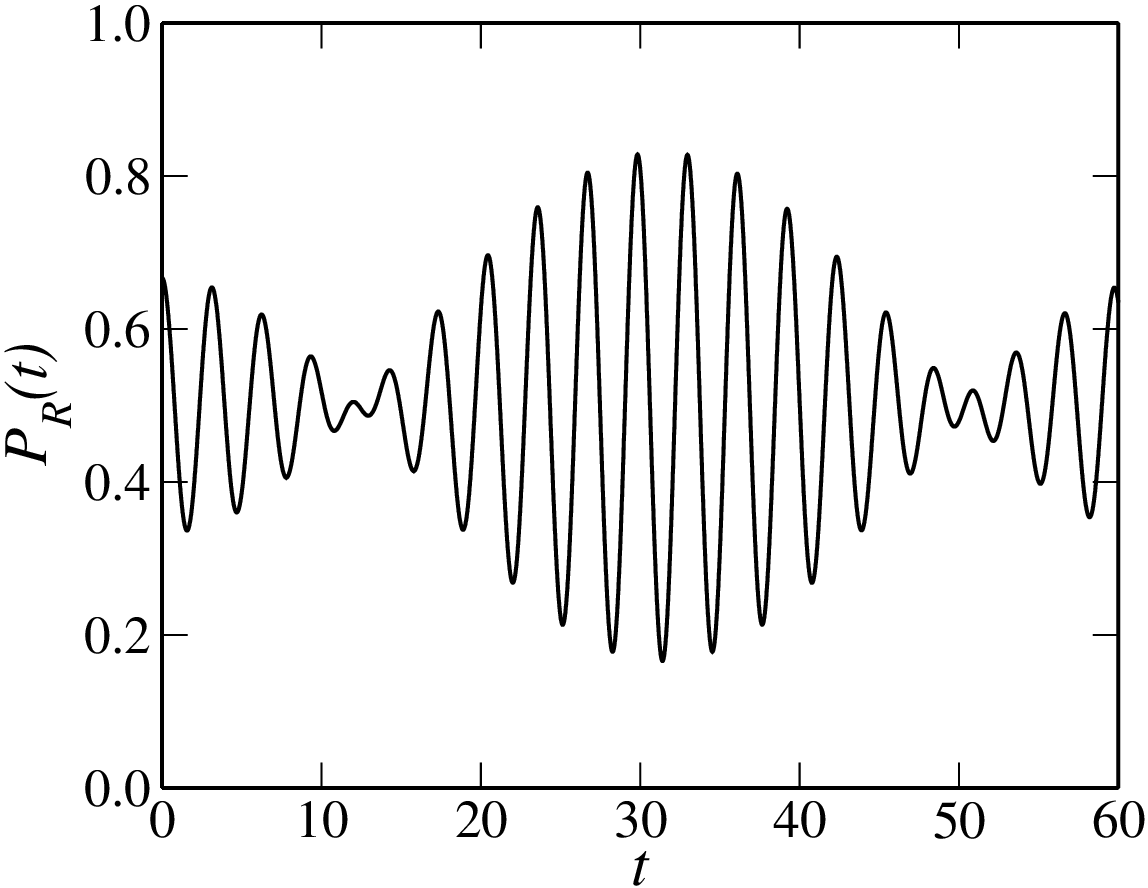}
\centerline{$(a)$ \hfill\hfill $(b)$ \hfill\hfill $(c)$ \hfill\hfill  $(d)$ \hfill\hfill}
\caption{\label{Three bosons}(Color online) $P_{R}(t)$ with the initial condition
$(1000)$ for three bosons. In $(a)$ square with dots (black)
denote $U'=0.05$ and solid line (red) denotes $U'=-0.1$ and $(b)$ 
solid line (black) denotes $U'=6$ and dotted line (red) denotes $U'=-12$. 
$P_{R}(t)$ with $(0100)$ initial
condition is depicted in 
$(c)$ with $U'=0.05$ and $(d)$
with $U'=-0.1$.}
\end{figure*}

     A close scrutiny of different initial conditions for three bosons in
the small $U$ regime reveals an interesting observation. $P_{R}(t)$,
corresponding to the initial condition $(0100)$ ($(0010)$), starts
with $\frac{2}{3}$ ($\frac{1}{3}$), as expected. However, as time
progresses, $P_{R}(t)$  modulates between values approximately
$0.85$ and $0.15$ (Fig.4(c) and (d)). Thus in the weak coupling regime ($U\simeq0.1$),
the fraction of the total number of bosons occupying the right well
is becoming larger (smaller) than $\frac{2}{3}$ ($\frac{1}{3}$),
thereby indicating a tendency of accumulation of particles in one of
the wells. This seems like an interesting result as an accumulation
of particles is not expected for $U<J$ ($U=0.1$ in units of $J$
here). $P_{R}(t)$ oscillating between values such as,  $\frac{2}{3}$
and $\frac{1}{3}$ may have been more commonly expected. The other initial condition, namely,
(1000) (or (0001)) does not exhibit any noteworthy feature and hence not included
for discussion.

    Therefore, it indicates that in case of odd number of
particles, when both the wells contain unequal number of particles
and the population difference between the wells differs by unity,
then such a scenario of accumulation of particles may be observed.
However in the noninteracting limit (with $V = U$), such
accumulation of particles vanishes and $P_{R}(t)$ oscillates between
$\frac{2}{3}$ and $\frac{1}{3}$. Similar result emerges for large $U$ limit ($U = 12$) 
where the accumulation of particles ceases to be a possibility owing to
trapping effects. We have skipped these plots for brevity.

      As an extension of the ongoing discussion, we take a look at the case of four bosons.
Here $P_{R}(t)$ is defined as,
\begin{equation}
\label{maa} P_{R}(t)=|c_{0}(t)|^2+\frac{3}{4}|c_{1}(t)|^2
+\frac{1}{2}|c_{2}(t)|^2 +\frac{1}{4}|c_{3}(t)|^2,
\end{equation}
 There is no qualitative difference in the behavior for $P_{R}(t)$ both in
$V = 2U$ and $V = 0.5U$ cases corresponding to
 the weak coupling regime between this and those for two or three
 bosons. In the strong  coupling case as expected, the localization
 is strong, and a complete tunneling of all the particles is
 prohibited over a very large time scales. Thus the notion of (Rabi)
 oscillations at large $U$ as inferred earlier, is no longer
 observed, at least for time scales $\sim 10^{3}$ (in limits of the
 tunneling frequency).

   Again an inspection of $P_{R}(t)$ with
 different initial conditions such as $(10000), (00001)$ and
 $(00100)$ yield results similar to those corresponding to $(100), (001)$ and
 $(010)$, respectively, for two bosons with the last one in either case
 yields, $P_{R}(t)=0.5$ for all $t$ and this value is fairly insensitive to the values of $U$ and $V$.
 For $(01000)$ and $(00010)$ as possible initial conditions, that is,
 three particle in the right well and one in the left, $P_{R}(t)$ oscillate
 between values $\frac{3}{4}$ and $\frac{1}{4}$, and vice versa as
 expected.

        Thus the effect of initial conditions seem to be important
 for an odd number of bosons in a double well, specially when
 the population difference between the well is unity. The claim is substantiated by
 looking at the case of five bosons, for which, at small
 values of $U$, three particles in the right well and two in the
 left (or vice versa) produces probabilities nearly $\frac{4}{5}$ and $\frac{1}{5}$ as time
 progresses, which are greater than $\frac{3}{5}$ and $\frac{2}{5}$, thereby indicating a
 possibility of accumulation of particles. However, four
 particles in one well and one in the other demonstrates no such
 accumulation tendencies, where $P_{R}(t)$ values oscillate between
 $\frac{4}{5}$ and $\frac{1}{5}$ as expected. The plots are skipped here for
 brevity.

\begin{figure*}[!t]
\centering\includegraphics[width=4.25cm]{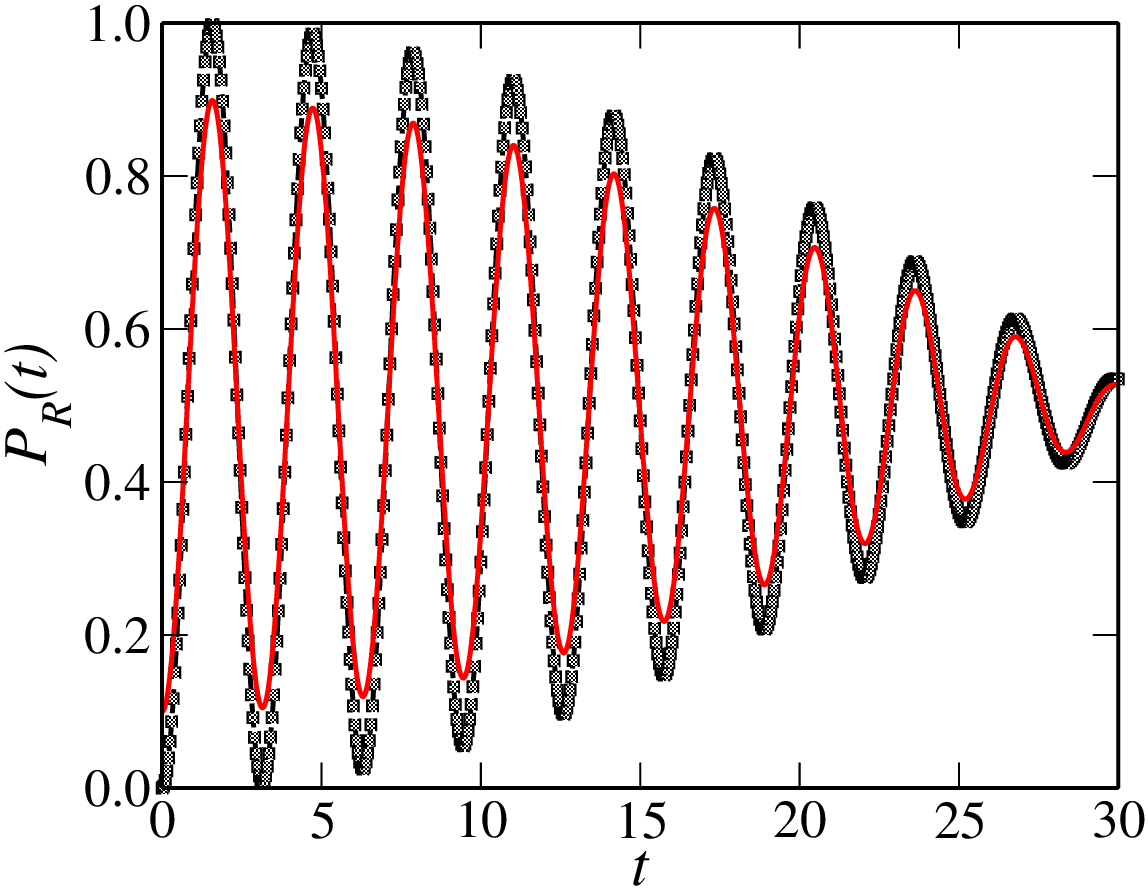}
\centering\includegraphics[width=4.25cm]{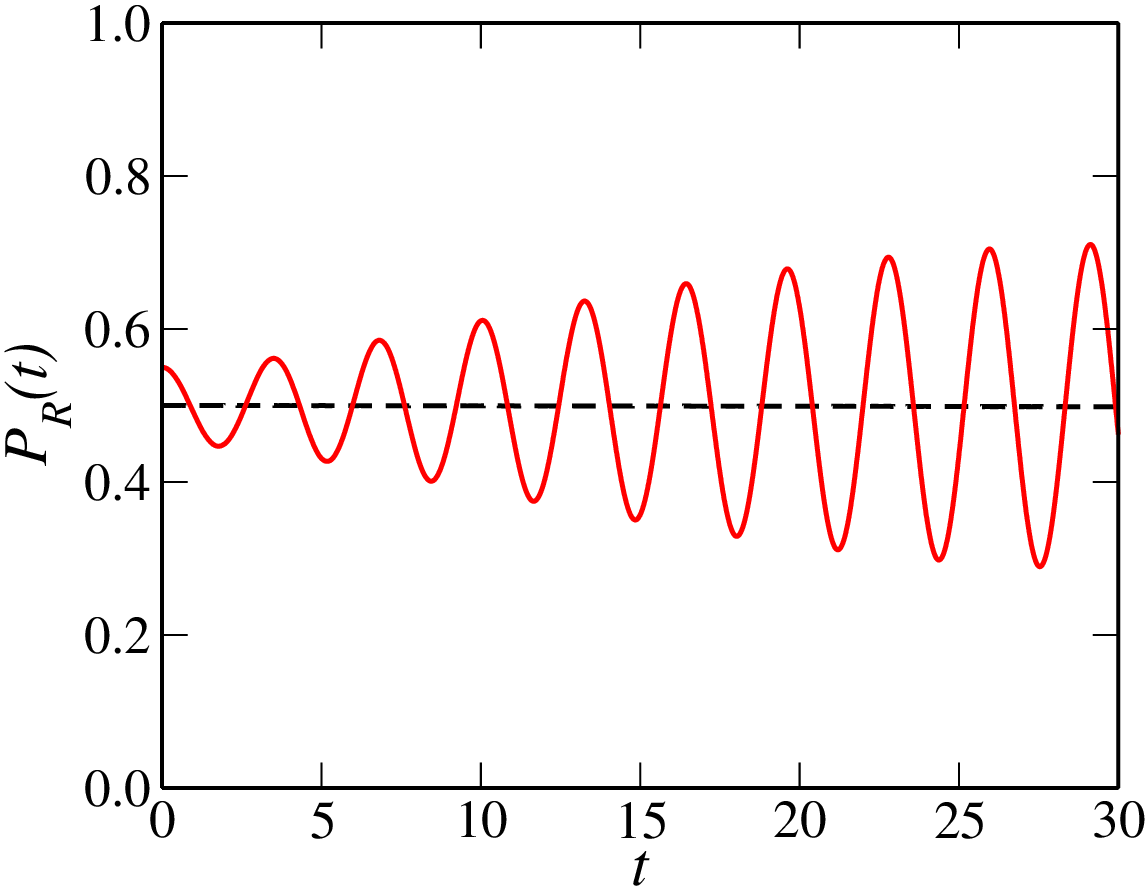}
\centering\includegraphics[width=4.25cm]{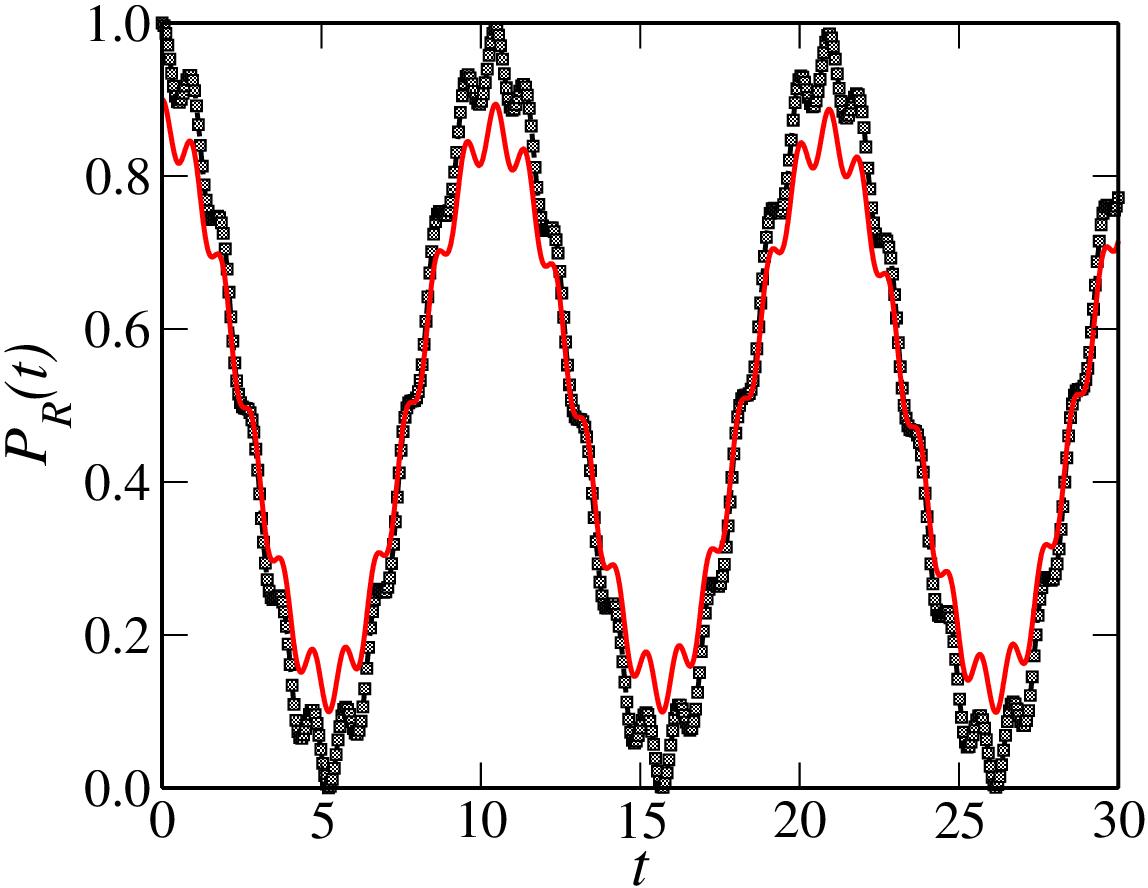}
\centering\includegraphics[width=4.25cm]{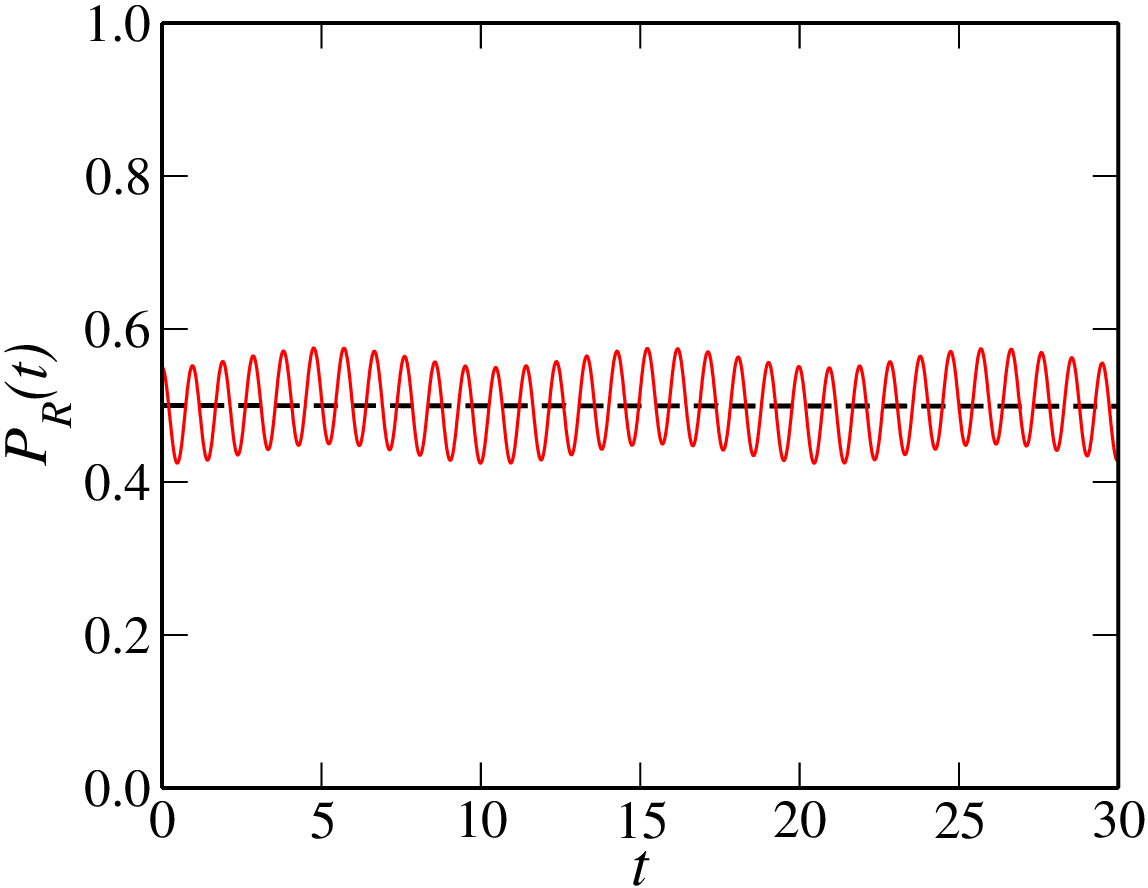}
\centerline{$(a)$ \hfill\hfill $(b)$ \hfill\hfill\qquad $(c)$ \hfill\hfill\qquad  $(d)$ \hfill\hfill}
\caption{\label{Admixture}(Color online) $P_{R}(t)$ with different
admixtures of initial conditions are shown for two bosons. In $(a)$
$U'=-0.1$, squares with dashed lines (black) denote the initial condition $(100)$
and solid line (red) denotes the initial condition $(\sqrt{0.9}, 0, \sqrt{0.1})$, 
$(b)$ $U'=-0.1$, dashed lines (black) denote the
initial condition $(010)$  and solid line (red) denotes the initial condition
$(\sqrt{0.1}, \sqrt{0.9}, 0)$, $(c)$ $U'=6$, 
squares with dashed lines (black) denote the initial condition $(100)$ and solid line (red) denotes the
initial condition $(\sqrt{0.9}, 0, \sqrt{0.1})$, $(d)$ $U'=6$ 
here dashed line (black) denotes the initial condition $(010)$  and solid line (red) denotes the initial condition
$(\sqrt{0.1}, \sqrt{0.9}, 0)$ A small admixture leads to an 
oscillatory dynamics in (b) and (d).}
\end{figure*}

\section{Admixture of states}

  Further emphasis on the effect of initial conditions can be given as follows. Instead of choosing a
particular initial state, one can consider an admixture of states. For example, for the 
case of two particles, instead of assigning an initial state (100), we may consider an admixture
of the form,
\begin{displaymath}
\beta (100) + \gamma (010) + \delta (001)
\end{displaymath} 
with the restriction, $|\beta|^{2} + |\gamma|^{2} + |\delta|^{2} = 1$. In this spirit, we have considered 
small deviations from the pure states by suitably choosing $\beta$, $\gamma$ and $\delta$ and
looked at the time evolved states via $P_{R}(t)$ for a comparison with those for the pure states.
The corresponding plots for some specific choices of $\beta$, $\gamma$ and $\delta$ 
are presented in Fig.(5). A slight deviation from a pure state, say (100) as
an initial state, results in a slightly different (Figs. 5 (a) and (c)) or completely
different dynamics (Figs.5 (b) and (d)) where in the latter case, a small mixing 
of the probability amplitudes corresponding to
(010) state yields an oscillatory dynamics. Thus the probability amplitudes of the initial 
states can slightly be modified to yield a desired oscillatory dynamics.


\section{Time averaged dynamics - extrapolation to large $N$}

In order to draw relevance of these results to the experiments done on cold atoms, we
need to extend the studies for a large number of bosons. An exact computation
of the quantum dynamics for such a large system is difficult. So we present a
time averaged $P_{R}(t)$, denoted by $\alpha$, which is defined as,
\begin{equation}
\alpha = \frac{1}{T} \int^{T}_{0} P_{R}(t)
\end{equation}
\begin{figure*}[!t]
\includegraphics[width=8.75cm]{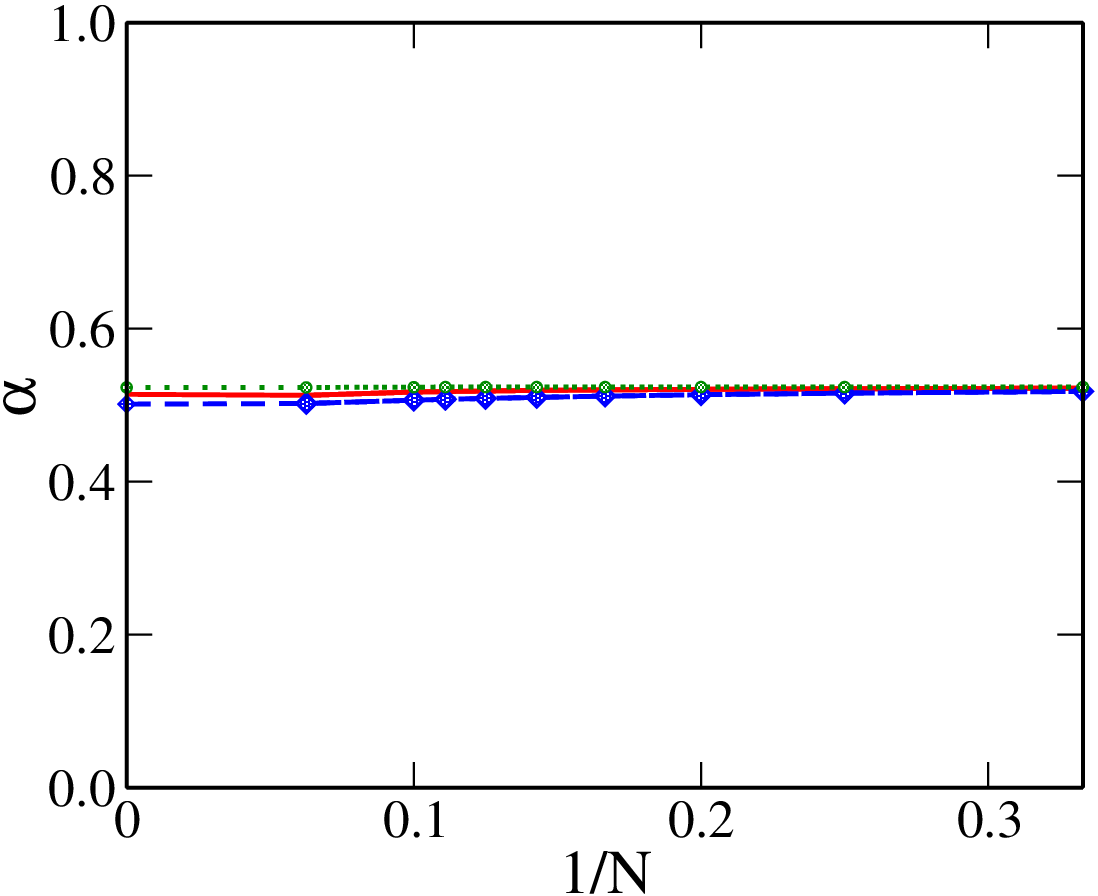}
\centering\includegraphics[width=8.75cm]{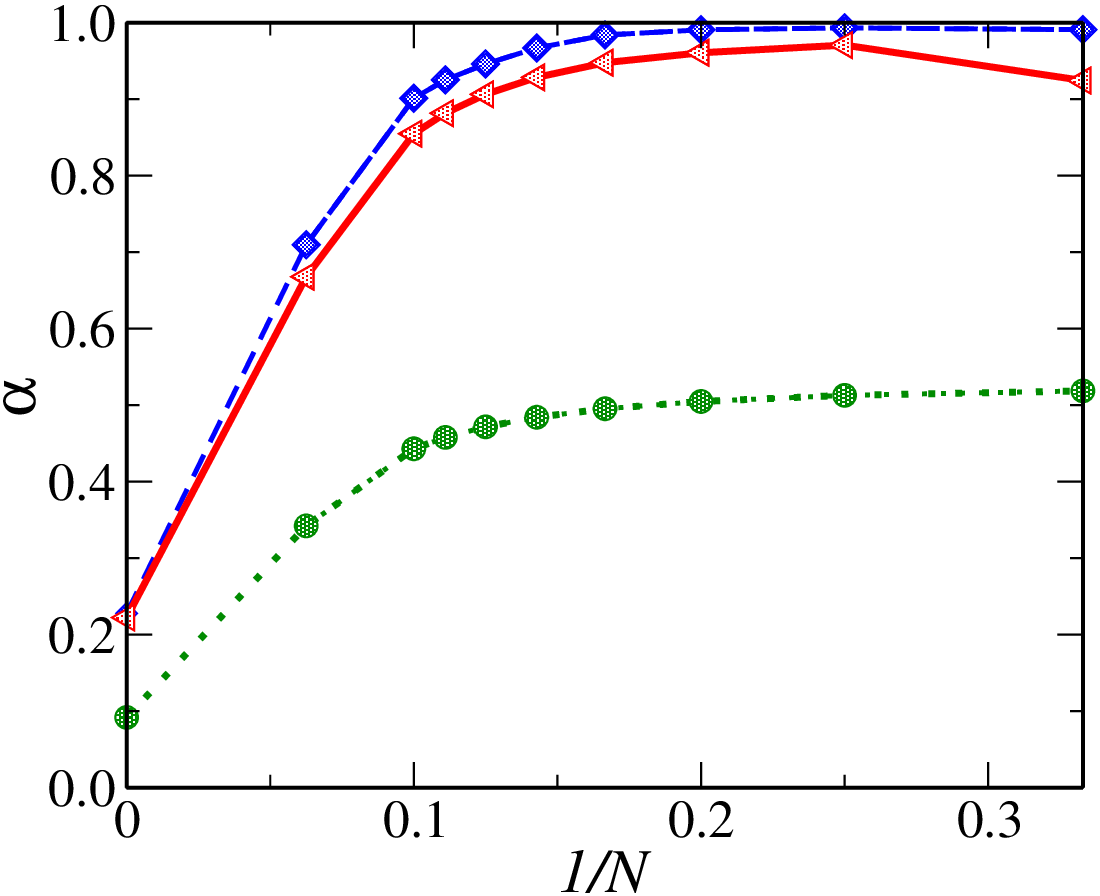}
\centerline{$(a)$ \hfill $(b)$ \hfill}
\caption{\label{Alpha}(Color online) Time integrated value of
$P_{R}(t)$ i.e. $\alpha$ as a function of inverse of the number of
bosons is shown for $(a)$ weak and $(b)$ strong coupling cases. In $(a)$ circles with dotted line (green) denotes
$\alpha$ for $V = U$ (non-interacting),solid line (red) denotes $\alpha$ for
$V=0.5U$ and diamonds with dashed line denotes $\alpha$ for $V=2U$. In $(b)$  circles with dotted line (green) denotes
$\alpha$ for $V = U$ (non-interacting), right triangles with solid line (red) denotes
$\alpha$ for $V=0.5U$ and diamonds with dashed line (blue) denotes $\alpha$ for $V=2U$. }
\end{figure*}
by computing the EOM exactly for upto 16 bosons and plotted $\alpha$ as a function of $1/N$ ($N$ being the
number of bosons) and hence the results are extrapolated to $N \rightarrow \infty$
(or $1/N \rightarrow 0$). Here $T$ is taken as $30$ in units of $1/J$ ($J$ being the 
tunneling amplitude). The results in the weak and strong coupling limits are presented in Fig.(6).
In Fig.6(a) which corresponds to a small $U$ regime, $\alpha$   does not depend on $N$ and 
stays at an average value of $0.5$, regardless of whether interaction effects have been included.
         In the large $U$ regime (in fig. Fig.6(b) ), $\alpha$ , although flat, yet different for
 different values of $V$ corresponding to smaller number of particles, shows a linear fall off as $N$ 
 becomes large. In the limit $N \rightarrow \infty$, $\alpha$ becomes small $0.1-0.2$ (the extrapolated value),
 re emphasizing the onset of the trapping effects as the time averaged probability for
 the particles to spend in one of the wells (right well here) becomes low. Hence there is indeed a depreciation in the value of the time averaged right well population in presence of a large number of bosons in a double well potential, however qualitatively similar physics can be expected as that for a few bosons.
        Summarizing the above discussion, we conclude by saying that the tunneling period not only increases
 with the particle number, but also depends on the interaction strength. Saturation behaviour of the time
 period  (that is intense trapping) is expected in the limit of N for the strongly interacting regime (which is clear from Fig.(6)).



\section{Conclusions}
We have carried out a detailed enumeration, though by no means exhaustive, of the effects of onsite
inter-particle repulsion on the tunneling dynamics of a few bosons
in a double well potential. The strong and weak coupling limits 
are compared and contrasted with regard to the study 
of tunneling dynamics.  Further, the sensitivity of the particle dynamics to different
initial conditions is closely scrutinized. For an odd number of
particles in the limit of weak repulsion, a population difference of
one particle among the two wells seems to demonstrate accumulation
tendencies. However, no such behavior is observed for the population
difference to be larger than one. Also the effect of an admixture of initial states 
on the tunneling oscillations has been studied. 
It is premature to comment on the
implication of this result to more elegant phenomena, such as using
it as an `\textit{atom switch}' etc, however we feel that our
results can motivate further experiments in the study of atomic
dynamics in presence of correlation effects.
\section*{Acknowledgement}
SB and SD thank CSIR, India for financial support under the grant -
F.No:03(1213)/12/EMR-II. AK acknowledges the useful discussion with L. Salasnich and 
support of IISER-Kolkata.
\section*{Author Contribution}
All authors contributed equally.


\begin{thebibliography}{99}
\bibitem{Anderson} M. H. Anderson, J. R. Ensher, M. R. Matthews, C.
E. Wieman, E. A. Cornell, Science {\bf{269}}, 198 (1995).
\bibitem{Davis} K. B. Davis, M. -O. Mewes, M. R. Andrews, N. J. van
Druten, D. S. Durfee, D. M. Kurn and W. Ketterle, Phys. Rev. Lett.
{\bf{75}}, 3969 (1995).
\bibitem{marco} 
A. N. Salgueiro, A. F. R. de Toledo Piza, G. B. Lemos, R. Drumond, M. C. Nemes, M. Weidemueller, Eur. Phys. J. D {\bf 44}, 537 (2007);
B. Juli\'{a}-D\'{\i}az, D. Dagnino, M. Lewenstein, J. Martorell, A. Polls, Phys. Rev. A {\bf 81}, 023615 (2010);
Q. Zhu, Q. Zhang and B. Wu, J. Phys. B: At. Mol. Opt. Phys. {\bf 48}, 045301 (2015);
B. Juli\'{a}-D\'{\i}az, J. Martorell, and A. Polls, Phys. Rev. A {\bf 81}, 063625 (2010);
M Mel\'{e}-Messeguer,5, B Juli\'{a}-D\'{\i}az, M Guilleumas, A Polls and A Sanpera, New J. Phys. {\bf 13}, 033012 (2011);
R. Lü, M. Zhang, J. L. Zhu, and L. You, Phys. Rev. A {\bf 78}, 011605(R) (2008);
M. A. Cazalilla, R. Citro, T. Giamarchi, E. Orignac, and M. Rigol, Rev. Mod. Phys. {\bf 83}, 1405 (2011);
B. Chatterjee, I. Brouzos, L. Cao, and P. Schmelcher, Phys. Rev. A {\bf 85}, 013611 (2012);
L. Cao, I. Brouzos, S. Z\"{o}llner and P. Schmelcher, New J. Phys. {\bf 13}, 033032 (2011);
L. Cao, I. Brouzos, B. Chatterjee and P. Schmelcher, New J. Phys. {\bf 14}, 093011 (2012);
\bibitem{Foling} S. F\"{o}lling, S. Trotzky, P. Cheinet, M. Feld, R. Saers, A. Widera,
T. M\"{u}ller, I. Bloch, Nature {\bf{448}}, 1029 (2007).
\bibitem{Levy} S. Levy, E. Lahoud, I. Shomroni and J. Steinhauer, Nature {\bf{449}},
579 (2007).
\bibitem{Albiez} M. Albiez, R. Gati, J. F\"{o}lling, S. Hunsmann, M. Cristiani, M. K. Oberthaler, Phys. Rev. Lett.
{\bf{95}}, 010402 (2005).
\bibitem{crespi} A. Crespi {\it et. al}, Nat. Photonics {\bf 7}, 545 (2013)
\bibitem{spring} J. B. Spring {\it et. al}, Science {\bf 339}, 798 (2013).
\bibitem{broome} M. A. Broome {\it et. al}, Science {\bf 339}, 794 (2013).
\bibitem{Wu} B. Wu and Q. Niu, Phys. Rev. A {\bf{61}} 023402 (2000).
\bibitem{Luo} X. Luo, Q. Xie and B. Wu, Phys. Rev. A {\bf{77}}, 053601
(2008).
\bibitem{Haroutyunyan} H. L. Haroutyunyan and G. Nienhuis, Phys. Rev. A {\bf{70}}, 063603
(2004).
\bibitem{Lignier} H. Lignier, C. Sias, D. Ciampini, Y. Singh, A.
Zenesini, O. Morsch and E. Arimondo, Phys. Rev. Lett. {\bf{99}},
220403 (2007).
\bibitem{Gong} J. Gong, L. M. -Molina and P. H\"{a}nggi, Phy. Rev. Lett. {\bf{103}},
133002 (2009).
\bibitem{Eckardt} A. Eckardt, M. Holthaus, H. Lignier, A. Zenesini, D. Ciampini, O. Morsch and E. Arimondo, Phys. Rev. A {\bf{79}}, 013611
(2009).
\bibitem{Gati} R. Gati and M. K. Oberthaler, J.Phys. B: At. Mol.
Opt. Phys. {\bf{40}}, R61 (2007).
\bibitem{Longhi} S. Longhi, J. Phys. B: At. Mol. Opt. Phys.
{\bf{44}}, 051001 (2011).
\bibitem{Smerzi} A. Smerzi, S. Fantoni, S. Giovanazzi and S. R.
Shenoy, Phys. Rev. Lett. {\bf{79}}, 4950 (1997).
\bibitem{Iskin} M. Iskin, Phys. Rev. A {\bf{83}}, 051606 (R) (2011).
\bibitem{Zollner} S. Z\"{o}llner, H. -D. Meyer and P. Schmelcher,
Phys. Rev. Lett. {\bf{100}}, 040401 (2008).
\bibitem{Stuhler} J. Stuhler, A. Griesmaier, J. Werner, T. Koch, M. Fattori and T. Pfau, J. Mod. Opt. {\bf{54}}, 647 (2007);
A. Gernier, J. Sebastian, P. Rehme, A. Aghajani Talesh, A
Griesmaier, T. Pfau, J. Phys. B {\bf{40}}, F77 (2007); A.
A. Griesmaier, J. Werner, S. Hensler, J. Stuhler and T.
Pfau, Phys. Rev. Lett. {\bf{94}}, 160401 (2005); P. O. Schimdt, S.
Hensler, J. Werner, A. Griesmaier, A. G\"{o}rlitz, T. Pfau and A.
Simoni, Phys. Rev. Lett. {\bf{91}}, 193201 (2003).
\bibitem{Buonsante} P. Bounsante, R. Buironi, E. Vescovi and A. Vezzani, Phys. Rev. A {\bf {85}}, 043625 (2012) 
\bibitem{Kolovsky} A.R. Kolovsky, J. Link and S. Wimberger, New J. Phys. {\bf {14}}, 075002 (2012)
\bibitem{Winkler} K. Winkler, G. Thalhammer, F. Lang, R. Grimm, J.
H. Denschlag, A. J. Daley, A. Kantian, H. P. B\"{u}chler and P.
Zoller,  Nature {\bf{441}}, 853 (2006).
\bibitem{Scully}  M. O. Scully and M. S. Zubairy in {\textit{Quantum Optics}}, Cambridge University Press  (1997)
\bibitem{Sachdev} S. Sachdev, K. Sengupta and S. M. Girvin, Phys. Rev. B {\bf{66}}, 075128 (2002).
\bibitem{Frazer} D. R. Dounas-Frazer, A. M. Hermundstad and L. D.
Carr, Phys. Rev. Lett. {\bf{99}}, 200402 (2007).
\end{thebibliography}
\end{document}